\documentstyle[aps,preprint]{revtex}

\begin{document}

\draft
\title{Direct subsurface absorption of hydrogen on Pd(111).}
\author{Ole Martin L{\o}vvik and Roar Aspes{\ae}ter
  Olsen\thanks{Present address: Theoretische Chemie, Vrije
    Universiteit, De Boelelaan 1083, 1081 HV Amsterdam, The
    Netherlands}}
\address{University of Oslo, Department of Physics, P.O.Box 1048
  Blindern, N--0316 Oslo, Norway}
\date{\today}
\maketitle

\begin{abstract}
  We summarize and discuss some of the available experimental and
  theoretical data important for understanding the role played by
  subsurface sites in dissociative chemisorption calculations for the
  H$_2$/Pd(111) system. Then we use a semi-empirical potential energy
  surface (PES) to model the interaction of a H$_2$ molecule impinging
  on a Pd(111) surface.  The London-Eyring-Polanyi-Sato (LEPS)
  construction has been extended to make direct subsurface absorption
  possible.  A 2-dimensional wave packet calculation is used to find
  qualitative trends in the direct subsurface absorption and to reveal
  the time scales involved.  We suggest that a partial in-plane
  relaxation occurs for the slowest incoming particles, thus resulting
  in a higher direct subsurface absorption probability for low
  energies.
\end{abstract}


\pagebreak

\section{Introduction}
\label{sec.introduction}

Hydrogen adsorption on Pd surfaces has during the last two decades
been extensively studied for a number of reasons.  Palladium has the
ability to absorb large amounts of hydrogen in the bulk, and is thus
interesting as a model system for understanding a good hydrogen
storage\cite{alefeld3,christmann1}.  The surface reactivity is also
important for layered systems, where a thin Pd surface may be used to
enhance hydrogen adsorption at low temperatures \cite{zhdanov1}.  The
interaction between hydrogen and transition metal surfaces has further
importance for the understanding of various catalytic reactions and
for metal embrittlement.  Hydrogen adsorption on metal surfaces also
has interest in itself, due to the quantum behaviour of the light H
atoms.

In understanding the formation of a hydride phase it is important to
know how the hydrogen enters the metal and how it is transferred from
the surface to the bulk.  Therefore a number of studies has been
investigating the role played by subsurface sites. In Section
\ref{sec.subsurface} we summarize and discuss some of the experimental
and theoretical data related to subsurface sites on Pd. In Section
\ref{sec.PES} we use this to construct a model PES for H$_2$ impinging
on a Pd(111) surface. Section \ref{sec.dynamics} describes the
techniques used for the dynamical calculations, and a discussion of
the model PES and results follow in Section \ref{sec.discussion}.
Section \ref{sec.conclusion} concludes.

\section{Subsurface sites on P\lowercase{d}}
\label{sec.subsurface}

Subsurface occupation connected with surface reconstruction has been
observed for Pd(110) for coverages ($\Theta$) larger than 1 monolayer
(ML) and at low temperatures (100--200 K)\cite{behm2,rieder2}.
Besenbacher {\em et al.\/}\cite{besenbacher1} found a saturation
coverage of 1 ML on the Pd(100) surface in the same temperature regime
when considering deuterium adsorption. They also noted that the
results of Ref.\ \cite{behm1} are consistent with no hydrogen
occupying subsurface sites at low temperatures. The theoretical
calculations of Ref.\ \cite{stauffer1} support this. Refs.\ 
\cite{hennig1,wilke1,wilke2} also agree that at low temperatures, 1 ML
is the expected saturation coverage. But they also state that at
higher temperatures the subsurface sites play an important role in
understanding the adsorption energy's dependence on coverages above 1
ML.

In the case of Pd(111), Eberhardt {\em et al.\/}\cite{eberhardt1} were
the first to postulate occupation of subsurface sites. After adsorbing
hydrogen on a low temperature substrate and then heating it to room
temperature, they concluded that the warming up caused a phonon
assisted conversion of hydrogen from the surface sites to lower energy
sites; the subsurface sites. The experimental results of Felter {\it
  et al.\/}\cite{felter2} showed hydrogen to form two distinct phases
with $\sqrt{3}\times\sqrt{3}$R30${}^\circ$ symmetry and with low
order-disorder transition temperatures; 85 and 105 K.  To establish
agreement between the experimental results and their calculations with
the embedded atom method (EAM), they found that occupation of
subsurface sites was crucial. When allowing the lattice to relax, the
surface and subsurface sites were found to be very close in energy and
separated by a small barrier. They also found the energies to be
coverage dependent, with subsurface sites being lowest for coverages
larger than 0.1 ML. For the two ordered phases with $\Theta =1/3$ and
$2/3$ ML, all the hydrogen was expected to go subsurface at 0 K. At
the critical temperature for the order-disorder transition,
approximately 60\% of the hydrogen occupied the subsurface sites for
both phases. These findings are consistent with electron-stimulated
desorption measurements of Kubiak and Stulen\cite{kubiak1}.

In Ref.\ \cite{daw1} Daw and Foiles gave a more detailed description
of their EAM results presented in Ref.\ \cite{felter2}. They found
only the following sites energetically favorable for hydrogen
occupation: surface tetrahedral (A$^+$), surface octahedral (B$^+$),
subsurface octahedral (A$^-$) and subsurface tetrahedral (B$^-$). The
classical energies, i.e.\ not including zero point energies, of these
sites were found to be from $-2.905$ to $-2.950$ eV.  This is in good
agreement with the experimental value for the adsorption energy
$-2.85$ eV of Conrad {\em et al.\/}\cite{conrad1} (the zero of the
energy scale is set to be that of a free hydrogen atom, and we have
neglected the zero point energies of the H$_2$ molecule and the
adsorbed H atom).  They again stressed that the energetical ordering
of the four sites was coverage dependent. Further they stated that
lattice relaxations are much more important for the B$^-$ sites
(relaxation energy 0.36 eV) than the A$^+$, A$^-$ and B$^+$ sites
(relaxation energies on the order of a few hundredths of an eV). The
calculated barrier height between the surface and subsurface sites for
low coverages was only 50 meV when including lattice relaxations. It
increased significantly with increasing coverage. For $\Theta =0.64$
ML they found only A$^-$ occupation at 0 K, 50\% subsurface occupation
above the critical temperature and almost no occupation of B$^-$ sites
at any temperature. Felter {\em et al.\/}, using
low-energy-electron-diffraction (LEED) measurements, returned to study
the $\Theta =2/3$ ML phase at 82 K in Ref.\ \cite{felter1} and found
the results to be consistent with a subsurface occupation in the range
0--60\%.

Cluster calculations of Rochefort {\em et al.\/}\cite{rochefort1}
found the A$^-$ sites to be 0.3--0.4 eV higher in energy than the
A$^+$ sites and the barrier height between the two sites to be 0.8 eV
(for an unrelaxed surface). Ezzehar {\em et al.\/}\cite{ezzehar1},
using a tight binding scheme not including lattice relaxations, found
the energy of the A$^+$ sites to be $-2.880$ eV, considerably lower
than the A$^-$ and B$^+$ sites ($-2.521$ and $-2.376$ eV,
respectively). This gave no support for subsurface occupation, but
they stated that this could be caused by the fact that they only
looked at isolated impurities. Rick {\em et al.\/}\cite{rick2} gave
some EAM values for the barrier height between surface and subsurface
sites. The A$^+$ and A$^-$ sites were separated by a 67 meV barrier,
B$^+$ and B$^-$ by 37 meV. These are the adiabatic values, i.e.\ 
allowing for lattice relaxations. The static lattice had 181 and 148
meV for the two barriers, respectively. Chakraborty {\em et
  al.\/}\cite{chakraborty1} also found the in-plane relaxations, i.e.\ 
relaxations of metal atoms within the top metal layer, to be very
important for the barrier height.  Recently Ezzehar {\em et
  al.\/}\cite{ezzehar2} pointed out a tendency to
$\sqrt{3}\times\sqrt{3}$R30${}^\circ$ superstructure formation with
hydrogen in both A$^+$ and A$^-$ sites, bringing them more in line
with the experiments of Felter {\em et al.\/}\cite{felter1}.

Also recently, L{\"o}ber and Hennig\cite{lober2} used the
full-potential linear muffin-tin orbital (FP-LMTO) method to study the
Pd(111) surface. They found the A$^+$ and B$^+$ sites (their notation
ffc or F and hcp or H, respectively) to be very close in energy,
$-2.67$ and $-2.69$ eV, for $\Theta=1$ ML. A slight coverage
dependence was found for the adsorption energies since for
$\Theta=1/3$ ML the respective energies were $-2.74$ and $-2.80$ eV.
These are in good agreement with the already mentioned experimental
value of $-2.85$ eV\cite{conrad1}. For the subsurface A$^-$ (their
notation O$_1$) sites they found the absorption energy to be $-2.60$
eV ($\Theta=1$ ML) when allowing for interlayer relaxations. The
barrier height for the A$^+$ to A$^-$ transition was calculated to
0.74 eV and found to be independent of interlayer relaxations. In
Ref.\ \cite{lober1} the barrier height was given as 0.82 eV under
slightly different considerations.

Thus for Pd(111), the following picture emerge:
\begin{itemize}
 \item There is good experimental and theoretical support for
       occupation of subsurface sites, even at low temperatures and
       low coverages.
 \item Interlayer relaxations are important for the depth of the
       subsurface well, but not for the barrier height.
 \item In-plane relaxations are very important for the barrier height
       between surface and subsurface sites.
\end{itemize}
However, this doesn't necessarily mean that the inclusion of
subsurface sites is important in a dissociative chemisorption
calculation. The population of subsurface sites could be a result of
thermally activated diffusion which is a slower process. We will come
back to this in Section \ref{sec.discussion}.

\section{The PES}
\label{sec.PES}

In the last years methods have been developed which, together with the
increasing computing power, now make it feasible to do {\em ab initio}
calculations of many points on the
PES\cite{hammer3,white1,wiesenekker1,wilke4}, and then use a set of
fit functions to interpolate between these values. These fit functions
are not necessarily chosen because of their simplicity, but because
they are flexible enough to allow a good fit to the {\em ab initio}
calculated points\cite{wiesenekker1,gross5}. There would be little
gained if the fit functions were chosen too simple to accommodate the
information of the {\em ab initio} data.

However, in the cases where the full PES is not available from {\em ab
  initio} calculations, it makes sense to choose a model potential
which is simple, yet reproduces the main expected features of the true
PES. Then one can use experimental data together with {\em ab initio}
calculated points to construct the model PES. The
London-Eyring-Polanyi-Sato (LEPS) potential, first introduced in
dissociative chemisorption by McCreery and Wolken\cite{mccreery2} and
later used by others\cite{mowrey4,cruz1,sheng1,dai3}, is well suited
to this. Fig.\ \ref{fig.interactions}a shows the three interactions
that go into the ordinary LEPS construction. Since we want to include
subsurface sites in our model PES, this form will not suffice.  We
therefore extend the LEPS formalism by introducing an additional
interaction representing the hydrogen interaction with the subsurface
and bulk metal layers as shown in Fig.\ \ref{fig.interactions}b. The
resulting three-dimensional model PES is then given by
        \begin{eqnarray}
        \label{eq.pes}
        V(r,z_1,z_2)&=&U_{\text{HH}}(r)+U_{\text{HM}}(z_1)+U_{\text{HM}}(z_2)\\
        &&+\sqrt{Q_{\text{HH}}(r)^2+(Q_{\text{HM}}(z_1)+Q_{\text{HM}}(z_2))^2-Q_{\text{HH}}(r)(Q_{\text{HM}}(z_1)+Q_{\text{HM}}(z_2))}\nonumber .
        \end{eqnarray}
Here $r$ is the hydrogen-hydrogen separation, and $z_1$ and $z_2$ are the
heights above/below the surface plane for the two hydrogen atoms
($z_1$ and $z_2$ take positive values above the plane, negative
below). The hydrogen-hydrogen (HH) interaction is described by
the two functions
        \begin{eqnarray}
        \label{eq.hydrogen}
        U_{\text{HH}}(r)&=& U(r,r_{0\text{HH}},\alpha_{\text{HH}},\Delta_{\text{HH}},D_{\text{HH}})\\
        Q_{\text{HH}}(r)&=& Q(r,r_{0\text{HH}},\alpha_{\text{HH}},\Delta_{\text{HH}},D_{\text{HH}}),\nonumber
        \end{eqnarray}
with the parameters  chosen to be those of a
hydrogen molecule far away from the surface (Table
\ref{tab:parameters}). The hydrogen-metal (HM) 
interactions are chosen as the  sum of the hydrogen-surface (HS)
interaction and the hydrogen-subsurface (HSS) interaction:
        \begin{eqnarray}
        \label{eq.metal}
        U_{\text{HM}}(z)&=& U(\mid z\mid,r_{0\text{HS}},\alpha_{\text{HS}},\Delta_{\text{HS}},D_{\text{HS}})+U(z+d_0,r_{0\text{HSS}},\alpha_{\text{HSS}},\Delta_{\text{HSS}},D_{\text{HSS}})\\
        Q_{\text{HM}}(z)&=& Q(\mid
z\mid,r_{0\text{HS}},\alpha_{\text{HS}},\Delta_{\text{HS}},D_{\text{HS}})+Q(z+d_0,r_{0\text{HSS}},\alpha_{\text{HSS}},\Delta_{\text{HSS}},D_{\text{HSS}}).
\nonumber
        \end{eqnarray}
The surface-subsurface spacing is  denoted $d_0$ and is equal to
4.24~$a_0$ for Pd(111).
The standard LEPS form follows for $U$
        \begin{equation}
        \label{eq.U}
        U(r,r_0,\alpha ,\Delta ,D)=\frac{D}{4(1+ \Delta )}\left(
        (3+\Delta )\exp\{-2 \alpha (r-r_0)\}-(2+6 \Delta )\exp \{- 
          \alpha (r-r_0)\}\right)
        \end{equation}
and $Q$
        \begin{equation}
        \label{eq.Q}
        Q(r,r_0,\alpha ,\Delta ,D)=\frac{D}{4(1+ \Delta )}\left(
        (1+3\Delta )\exp\{-2 \alpha (r-r_0)\}-(6+2 \Delta )\exp \{-
          \alpha (r-r_0)\}\right) .
        \end{equation}
The parameters for the hydrogen-metal interactions in Table
\ref{tab:parameters} are chosen to fit values for atomic hydrogen
approaching the Pd(111) surface above the surface tetrahedral (A$^+$)
site, and going down to the subsurface octahedral (A$^-$) site. We
have constructed two different PESes 1 and 2 with barrier heights
$E_{\text{B}}=0.8$ and 0.4 eV per H atom, respectively. The motivation
for this is given in Section \ref{sec.discussion}. Both PESes mimic
the H$_2$ molecule approaching the surface with the molecular axis
parallel to the plane of the surface and dissociating above a bridge
site into the threefold hollow sites. This is a favorable adsorption
geometry and also allows the hydrogen atoms to go to the subsurface
sites (A$^-$, B$^-$) directly below the threefold surface sites
(A$^+$, B$^+$). For simplicity we have assumed the same energetics for
the B and A sites. In the dynamical calculations we only have two
degrees of freedom: $r$ and $Z$, the height of the center of mass
above the surface. This means that we are restricted to that both
atoms in the molecule either go to the surface or to the subsurface.
Fig.\ \ref{fig.subsurface} shows a contour plot of the resulting PES
1, given by $V(r,Z,Z)$.

\section{dynamical calculations}
\label{sec.dynamics}

The two-dimensional Hamiltonian operator governing the motion of a
hydrogen molecule impinging on a flat, rigid metal surface is
        \begin{equation}
        \label{eq.hamilton}
        \hat{H}= \frac{1}{2M} \frac{\partial^2}{\partial Z^2} +
                 \frac{1}{2 \mu} \frac{\partial^2}{\partial r^2} + V(r,Z,Z) .
        \end{equation}
We use atomic units unless otherwise explicitly stated, which gives
the total mass $M=3674.4$ and the reduced mass $\mu=918.6$. $V(r,Z,Z)$ is
the model PES described in the previous section. The 
time independent Hamiltonian gives the formal solution to the
Schr{\"o}dinger equation
        \begin{equation}
        \label{eq.schroedinger}
        \Psi(r,Z,t+\delta t)= e^{-i \hat{H} \delta t} \: \Psi(r,Z,t) .
        \end{equation}
We use a $34.5$ by $15.0$ grid, with $Z$ ranging from $-6.5$ to $28.0$
and $r$ from 0 to $15.0$. 512 and 64 points are used in the
$Z$ and $r$ directions, respectively. The numerical time evolution of the
wave function is obtained by expanding the time evolution operator
$e^{-i \hat{H} \delta t}$ according to the Chebychev
technique\cite{tal-ezer1}. The action of the potential energy operator
on the wave function is found by multiplying the wave function by the
potential energy at each grid point. Added to this is the kinetic
energy part of the Hamiltonian, obtained by the FFT technique of
Kosloff and Kosloff\cite{kosloff2}.  The initial wave function is
chosen to be a product of the vibrational ground state of the hydrogen
molecule, $\chi_0(r)$, and a Gaussian in $Z$:
        \begin{equation}
        \label{eq.}
        \Psi(r,Z,t_0)= \chi_0(r) \: (\pi \delta^2)^{-1/4} \:
                       exp\left\{-\frac{(Z-Z_0)^2}{2 \delta^2} +
                             i k_0 Z \right\} .
        \end{equation}

The shape of the vibrational ground state wave function was obtained
analytically from Ref.\ \cite{pekeris1}, whereas the normalisation constant is
found numerically.  $Z_0$ and $k_0$ define the initial position and
momentum, respectively.  For all the runs we start the wave packet at
$Z_0=15.0$. This far away from the surface the PES has obtained its
asymptotic value in the $Z$ direction. Choosing $Z_0$ smaller would cause the
wave packet to miss a part of the drop in potential energy towards the
surface, thus reducing the kinetic energy with which the wave packet is
approaching the barrier.

Since we have not performed an asymptotic analysis of the wave packet
in line with Refs.\ \cite{balint-kurti1,balint-kurti2,mowrey2}, we
want to have a narrow energy distribution in our initial wave packet,
thus $\delta=3.0$. We will then take $E_0=k_0^2/2M$ to be the initial
kinetic energy of the wave packet. To avoid problems with the wave
function at the boundary of the grid, we multiply it by an exponential
damping function after each time step. The damping function is 1 in
the entrance channel and the interaction region. In the exit channel
it is continuously decreasing from 1 to 0 towards the $r=15.0$
boundary of the grid. The direct subsurface absorption probability,
$\eta_{\text{SS}}$, is found by accumulating the removed norm for $Z$
less than 0. Norm conservation is checked.

The Chebychev method is known to be most efficient for long time
steps\cite{tal-ezer1}. But since we only damp the wave function after
each time step (as opposed to the use of optical potentials, where the
damping function effectively is a part of the
Hamiltonian\cite{neuhauser1,vibok1,vibok2}), a too long time step
would allow the wave function to reach the boundary of the grid. We
have found that a time step of 300~a.~u.\ is short enough to avoid
this problem, yet long enough for the Chebychev technique to be
efficient. Repeated runs with different initial kinetic energies then
give the energy dependence of $\eta_{\text{SS}}$.

\section{Discussion}
\label{sec.discussion}

The simple form of our model PES carries some limitations.  Even
though there are quite a few tunable parameters in our model PES, the
Morse form of $U$ and $Q$ in Eqs.\ \ref{eq.U} and \ref{eq.Q} limits
how many points we can accurately fit. The positions of the surface
and subsurface minima, $r_{\text{S}}$ and $r_{\text{SS}}$, are bound
to be close in absolute value.  Felter {\em et al.\/}\cite{felter1}
gave values from $1.5$ to $1.6$ for $r_{\text{S}}$ and from $-2.2$ to
$-2.3$ for $r_{\text{SS}}$, whereas our model PES has
$r_{\text{S}}=1.5$ and $r_{\text{SS}}=-1.7$. Also the subsurface
absorption minimum ($E_{\text{SS}}$) always lies slightly below the
surface adsorption minimum ($E_{\text{S}}$). This fits in the case of
a relaxed lattice \cite{felter2,daw1,rick2}, but not for the static
lattice \cite{rochefort1,ezzehar1} where $E_{\text{SS}}$ is
significantly {\em higher\/} than $E_{\text{S}}$. However, we think
this is not too important for our dynamical calculations.  The most
important quantity in determining whether the hydrogen goes to the
surface or to the subsurface is the barrier height. The values of
$E_{\text{SS}}$ and $r_{\text{SS}}$ will be more important when
discussing thermally activated diffusion between surface and
subsurface sites. This process is however many orders of magnitude
slower than the dissociative adsorption/absorption process, as is
indicated by the low surface to subsurface rates of Rick {\em et
  al.\/}\cite{rick2}.

Another possible problem with our model PES is the cusp at $Z=0$
introduced by the dependence of $\mid Z\mid$ in Eq.\ \ref{eq.metal}.
It yields a barrier that is somewhat narrower than that of the
expected smooth physical PES. At the same time, we see that the
barrier height is somewhat larger near the entrance channel than in
the asymptotic region (large $r$ values). We have fitted our model PES
to values for atomic hydrogen interacting with a metal surface, and
the details of the interaction region are not yet known. Thus, our
barrier height should be taken {\em cum grano salis\/}. Having said
this, the increasing barrier height for small $r$ values is easy to
understand physically. For the hydrogen molecule to dissociate and
absorb, the atoms have to move towards the favorable absorption sites.
Fig.\ \ref{fig.snapshot} shows a snapshot of the wave function when it
is crossing the barrier, obtained from the dynamical calculations. We
can see that it crosses the barrier around $r=4$, about the same as
the distance between two neighboring hollow surface sites which the
atoms pass close to on the way subsurface. We therefore believe that
our model PES has the qualitatively correct features for describing
direct subsurface absorption.

In Section \ref{sec.subsurface} we mentioned the diverging results
when it comes to the values of the barrier heights.  For a static
lattice the calculated barrier heights vary from 148 and 181 meV
\cite{rick2} to 0.74 and 0.76 eV \cite{lober2}, 0.8 eV
\cite{rochefort1}, and 0.82 eV \cite{lober1} per H atom.  Clearly this
will have drastic effects on the probability of subsurface
absorption. L{\"o}ber \cite{lober1}, L{\"o}ber and Hennig
\cite{lober2}, and Rochefort {\em et al.\/}\cite{rochefort1} all work
within the local density approximation (LDA).  It has been shown by
others\cite{hammer3,white1,wiesenekker1} that LDA is not able to
reproduce the experimental barrier height of the H$_2$/Cu system.
This barrier is different from the one we are considering in the
respect that it is located in front of, and not within, the
surface. But if the trends of the H$_2$/Cu system is followed by the
H$_2$/Pd system, it means that $E_{\text{B}}=0.74$ eV \cite{lober2} is
too low. But L{\"o}ber and Hennig state that this value is already too
high when compared to available experimental data. This, in addition
to the large discrepancies between the LDA barrier heights
\cite{rochefort1,lober2,lober1} and the EAM values of Rick {\em et
al.\/}\cite{rick2}, shows that more effort is needed to determine a
more reliable value. For our model PES 1, we have chosen the
parameters to give a barrier height $E_{\text{B}}=0.8$ eV per H atom.

In Section \ref{sec.subsurface} we also saw that allowing for in-plane
relaxations gave much lower values for the barrier height. Therefore
it is important to answer the question: does the lattice have time to
relax?  The estimated surface Debye temperature of Pd is
$T_D=164$~K\cite{rick2}. This gives a measure of the maximum surface
phonon frequency $\omega_D$ and sets a typical time scale ($t_s$) for
the surface phonons to $t_s=2 \pi / \omega_D = 2 \pi / k_B
T_D=12000$~a.~u.  We suppose that this is about the same time
requested to perform the in-plane relaxation, since both the
relaxation length and the amplitude of the phonon vibration are
typically a few percent of $d_0$.  Next we have to know how much time
the wave packet spends in the barrier region. Studying snapshots of
the wave function like Fig.\ \ref{fig.snapshot} reveals that this time
varies from about 1000~a.~u.\ for the highest initial kinetic energies
to more than 5000~a.~u.\ for the lowest ones. Thus when $E_0$ is high
the wave packet spends a shorter time in the barrier region than it
would take the nearby substrate atoms to relax, thence it is
reasonable to use a barrier height representing a static lattice.
Nevertheless, since the time needed for relaxation is much shorter
than the diffusion time scale, the interlayer relaxed value for
$E_{\text{SS}}$ should be used when discussing diffusion back to the
surface.  For a lower $E_0$, the wave packet spends more time in the
barrier region and the nearby substrate atoms should have time to at
least partially relax. This will cause the barrier height to drop. The
longer the wave packet spends in the barrier region, the more the
surface has time to relax, and the larger this drop will be. The
parameters of our model PES 2 have been chosen to give a barrier
height $E_{\text{B}}=0.4$ eV per H atom, which we will think of as
that of a partially relaxed lattice valid for a small $E_0$.

With this low barrier height, the barrier actually lies below the
energy of the free hydrogen molecule, so most of the wave function is
expected to go directly to the subsurface.  If we had used even lower
values for the static lattice barrier height, the barrier would have
been so small compared to the kinetic energy that is gained on the way
towards the surface, that the wave function would be smeared out
between the surface and subsurface site even for small kinetic
energies.  It would not be localized before the thermalization
occurred.  At that time the relaxation also should have finished, so
that $E_{\text{S}}$ and $E_{\text{SS}}$ are almost degenerate.  With a
low barrier height, we could thus expect equal probabilities for
finding hydrogen at the surface and subsurface, independent of $E_0$.

Fig.\ \ref{fig.SS-probability} shows the results of the dynamical
calculations for the two PESes. For a static lattice (PES 1) the
subsurface gets appreciably populated only for $E_0$ above 0.7 -- 0.8
eV. The partially relaxed lattice, represented by PES 2, shows almost
70\% subsurface occupation already for the smallest initial kinetic
energies. If the lattice was allowed to relax during the dynamical
calculations, the resulting $\eta_{\text{SS}}$ is expected to be a
combination of the curves for PES 1 and PES 2. Without introducing a
time dependent PES and allowing for energy exchange between the
hydrogen molecule and substrate atoms, only qualitative trends can be
given. For low $E_0$, there are two competing effects. Firstly, an
increase in $E_0$ makes it easier for the wave packet to cross the
barrier.  Secondly, the nearby substrate atoms get a shorter time to
relax, so that the effective barrier height increases.  The net result
is a decrease in $\eta_{\text{SS}}$.  As we continue to raise the
initial kinetic energy, the lattice gets even shorter time to relax.
At a certain point, the increasing kinetic energy is more important
than the increasing barrier height, and $\eta_{\text{SS}}$ then starts
to grow.  This qualitative trend is suggested by the dashed line in
Fig.\ \ref{fig.SS-probability}.

As pointed out by the referee, also the barrier {\em width\/} will
change when the surface has time to relax.  This might affect the
behaviour of $\eta_{\text{SS}}$ for low energies, since the tunneling
wave packet has a complicated interaction in the intermediate region.
The size of this effect is difficult to estimate without going through
a full time dependent analysis.  However, we do not believe that the
barrier width will change substantially due to the in-plane
relaxation, since then the positions of the surface and subsurface
minima are fixed.  In Fig.\ \ref{fig.SS-probability} we can see a
small effect of the change in barrier width following a lower barrier
height: the slope of the curve for PES 2 is slightly steeper than for
PES 1 in the same region of $\eta_{\text{SS}}$.  Interplane
relaxations typically increases the interplane distance and hence the
barrier width with only a few percent, not enough to make any major
contributions to $\eta_{\text{SS}}$.  The upper energy limit should
moreover be unaffected, so we believe that the global trend will have
the same qualitative features as the suggested curve in Fig.\ 
\ref{fig.SS-probability}.

In Section \ref{sec.subsurface} we mentioned that the experimentally
observed occupation of subsurface sites could be a result of thermally
activated diffusion from surface sites. Even though our results are
dependent on the particular geometry we have chosen for the incoming
molecule, we have demonstrated the possibility for direct
subsurface absorption taking place on the same time scale as
dissociative adsorption.  This possibility is seen both for low and
high barrier heights, the only difference is in the dependency on
$E_0$.

\section{Conclusion}
\label{sec.conclusion}

The most important quantity when discussing direct subsurface
absorption is the barrier height between the surface and subsurface.
This is however not known experimentally, and the calculated values
vary a lot.  Calculations have further shown that in-plane relaxations
lowers the barrier height significantly, and it is therefore crucial
to decide whether the surface has enough time to relax.  Our dynamical
calculations have shown that high-energetic molecules may be viewed in
the static limit, that is with a high barrier height.  When the
hydrogen molecule has lower kinetic energy, however, the time spent in
the barrier region is comparable to the time needed to perform a
relaxation of the metal lattice. For a barrier height of 0.8~eV per H
atom, we suggest that this results in a probability of direct
subsurface absorption that is decreasing for low initial kinetic
energies due to the increasing barrier height.  For higher energies,
the barrier height stabilizes, and the probability of overcoming the
barrier thus increases.  For barrier heights below 0.4~eV per H atom,
we expect that about 50\%\ of the hydrogen is found subsurface after
thermalization, independent of the initial kinetic energy.

The phonon mode associated with the variation of the barrier height
has been identified by Chakraborty {\em et al.\/}\cite{chakraborty1}
and Rick {\em et al.\/}\cite{rick2}. With the total energy
calculations techniques now available, it should in a future study be
possible to calculate the frequency of this phonon excitation. This
would clarify the extent of the coupling between the dissociation
process and this phonon mode, and thus check the validity of using the
static limit.

Our results are of course dependent on the limitations of the
two-dimensional calculations and the particular geometry chosen.  To
obtain more reliable results it would be necessary to work with at
least three dimensions.  It would also be valuable to have more points
to the  fitting of  the PES.  We therefore believe that total energy
calculations to create a full three- or higher-dimensional PES in the
line of Refs.\ \cite{hammer3,white1,wiesenekker1,wilke4} should be
taken on.

\section{Acknowledgments}
\label{sec.acknowledgments}
We would like to thank Geert-Jan Kroes for helpful discussions and
suggestions on the dynamical calculations and Runa L{\"o}ber, Axel
Gross and Steffen Wilke for sending us their results prior to
publication. R.A.O.\ would like to thank George Darling and Stephen
Holloway for organizing the Dynamics Workshop in Chester, U.K., which
offered a very good opportunity to discuss the most recent ideas on
quantum dynamical calculations and potential energy surfaces. Many
thanks to the participants of this workshop as well, especially Rick
Mowrey for being so positive and encouraging. Finally, we would like
to thank John Rekstad for his support. This work has been
financed by the Norwegian Research Council.

\begin{figure}[p]
  \caption{The interactions used in a) the ordinary and b) our
    modified LEPS potential energy surface.  Solid, dashed and dotted
    lines specifies the hydrogen-surface, the hydrogen-hydrogen, and
    the hydrogen-subsurface interactions, respectively.}
  \label{fig.interactions}
\end{figure}

\begin{figure}[p]
\caption{The potential energy surface 1 for H$_2$ on Pd(111) used in
  the calculations.  The barrier height is $E_{\text{B}}=0.8$ eV per H
  atom, and the contour spacing is 0.5 eV.  The model parameters are
  listed in Table \protect\ref{tab:parameters}.}
\label{fig.subsurface}
\end{figure}

\begin{figure}[p]
\caption{A snapshot of the absolute square of the wave function taken
  at t=4200 a.~u.  The initial kinetic energy was 0.89 eV, the initial
  width in the $Z$ direction was $\delta$=3.0 $a_0$, and it started at
  $Z_0$=15.0 $a_0$.  The potential energy surface 1 with barrier height
  $E_{\text{B}}=0.8$ eV per H atom was used in the dynamics.}
\label{fig.snapshot}
\end{figure}

\begin{figure}[p]
\caption{The probability of direct subsurface absorption
  $\eta_{\text{SS}}$ as a function of initial kinetic energy $E_0$ for
  the static (PES 1) and partially relaxed (PES 2) lattice.  The
  dashed curve indicates qualitatively how we think the curve would go
  if the lattice was allowed to do in-plane relaxations during the
  dynamical calculations.}
\label{fig.SS-probability}
\end{figure}

\begin{table}[p]
  \begin{center}
    \leavevmode
    \begin{tabular}[c]{ldddd}
      ~ & $r_0$[$a_0$] & $ \alpha$ [$a_0^{-1}$] & $ \Delta $[1] &
      $D$[eV] \\
      \hline
      H-H& 1.4002 & 1.0282 & 0.05 & 4.745 \\
      H-S (PES 1)& 1.613 & 0.29 & 0.2 & 2.694 \\
      H-SS (PES 1)& 2.0 & 0.29 & 0.2 & 0.272  \\
      H-S (PES 2)& 1.613 & 0.214 & 0.2 & 2.656 \\
      H-SS (PES 2)& 2.0 & 0.214 & 0.2 & 0.272
    \end{tabular}
  \end{center}
  \caption{Parameters for the two different potential energy surfaces
    1 and 2 used in the dynamical calculations.  PES 1 gives a barrier
    height $E_{\text{B}}=0.8$ eV per H atom.  PES 2 has
    $E_{\text{B}}=0.4$ eV per H atom. Both surfaces have
    $E_{\text{S}}=-2.85$ eV per H atom, $E_{\text{SS}}=-2.95$ eV per H
    atom, $r_{\text{S}}=1.5$ $a_0$, and $r_{\text{SS}}=-1.7$ $a_0$.}
  \label{tab:parameters}
\end{table}


\begin{thebibliography}{10}

\bibitem{alefeld3}
{\em Hydrogen in metals}, edited by G. Alefeld and J. V{\"o}lkl (Springer
  Verlag, Berlin, 1978).

\bibitem{christmann1}
K. Christmann, Surf. Sci. Rep. {\bf 9},  1  (1988).

\bibitem{zhdanov1}
V.~P. Zhdanov, A. Krozer, and B. Kasemo, Phys. Rev. B {\bf 47},  11044  (1993).

\bibitem{behm2}
R.~J. Behm {\it et~al.}, J. Chem. Phys. {\bf 78},  7486  (1983).

\bibitem{rieder2}
K.~H. Rieder, M. Baumberger, and W. Stocker, Phys. Rev. Lett. {\bf 51},  1799
  (1983).

\bibitem{besenbacher1}
F. Besenbacher, I. Stensgaard, and K. Mortensen, Surf. Sci. {\bf 191},  288
  (1987).

\bibitem{behm1}
R.~J. Behm, K. Christmann, and G. Ertl, Surf. Sci. {\bf 99},  320  (1980).

\bibitem{stauffer1}
L. Stauffer, R. Riedinger, and H. Dreyss\'e, Surf. Sci. {\bf 238},  83  (1990).

\bibitem{hennig1}
D. Hennig, S. Wilke, R. L{\"o}ber, and M. Methfessel, Surf. Sci. {\bf 287/288},
   89  (1993).

\bibitem{wilke1}
S. Wilke {\it et~al.}, Surf. Sci. {\bf 307-309},  76  (1994).

\bibitem{wilke2}
S. Wilke, D. Hennig, and R. L{\"o}ber, Phys. Rev. B {\bf 50},  2548  (1994).

\bibitem{eberhardt1}
W. Eberhardt, F. Greuter, and E.~W. Plummer, Phys. Rev. Lett. {\bf 46},  1085
  (1981).

\bibitem{felter2}
T.~E. Felter, S.~M. Foiles, M.~S. Daw, and R.~H. Stulen, Surf. Sci. Lett. {\bf
  171},  L379  (1986).

\bibitem{kubiak1}
G.~D. Kubiak and R.~H. Stulen, J. Vac. Sci. Technol. A {\bf 4},  1427  (1986).

\bibitem{daw1}
M.~S. Daw and S.~M. Foiles, Phys. Rev. B {\bf 35},  2128  (1987).

\bibitem{conrad1}
H. Conrad, G. Ertl, and E.~E. Latta, Surf. Sci. {\bf 41},  435  (1974).

\bibitem{felter1}
T.~E. Felter, E.~C. Sowa, and M.~A. van Hove, Phys. Rev. B {\bf 40},  891
  (1989).

\bibitem{rochefort1}
A. Rochefort, J. Andzelm, N. Russo, and D.~R. Salahub, J. Am. Chem. Soc. {\bf
  112},  8239  (1990).

\bibitem{ezzehar1}
H. Ezzehar, L. Stauffer, and H. Dreyss{\'e}, Z. Phys. Chem. {\bf 181},  305
  (1993).

\bibitem{rick2}
S.~W. Rick, D.~L. Lynch, and J.~D. Doll, J. Chem. Phys. {\bf 99},  8183
  (1993).

\bibitem{chakraborty1}
B. Chakraborty, S. Holloway, and J.~K. N{\o}rskov, Surf. Sci. {\bf 152/153},
  660  (1985).

\bibitem{ezzehar2}
H. Ezzehar, L. Stauffer, H. Dreyss{\'e}, and M. Habar, Surf. Sci. {\bf
  331--333},  144  (1995).

\bibitem{lober2}
R. L{\"o}ber and D. Hennig, subm. to Phys. Rev. B.

\bibitem{lober1}
R. L\"ober, Ph.D. thesis, Humboldt-University of Berlin, 1995.

\bibitem{hammer3}
B. Hammer, M. Scheffler, K.~W. Jacobsen, and J.~K. N{\o}rskov, Phys. Rev. Lett.
  {\bf 73},  1400  (1994).

\bibitem{white1}
J.~A. White, D.~M. Bird, M.~C. Payne, and I. Stich, Phys. Rev. Lett. {\bf 73},
  1404  (1994).

\bibitem{wiesenekker1}
G. Wiesenekker, G.~J. Kroes, E.~J. Baerends, and R.~C. Mowrey, J. Chem. Phys.
  {\bf 102},  3873  (1995).

\bibitem{wilke4}
S. Wilke and M. Scheffler, subm. to Phys. Rev. B.

\bibitem{gross5}
A. Gross, S. Wilke, and M. Scheffler, Phys. Rev. Lett. {\bf 75},  2718  (1995).

\bibitem{mccreery2}
J.~H. McCreery and {G. Wolken, Jr.}, J. Chem. Phys. {\bf 63},  2340  (1975).

\bibitem{mowrey4}
R.~C. Mowrey, J. Chem. Phys. {\bf 94},  7098  (1991).

\bibitem{cruz1}
A.~J. Cruz and B. Jackson, J. Chem. Phys. {\bf 94},  5715  (1991).

\bibitem{sheng1}
J. Sheng and J.~Z.~H. Zhang, J. Chem. Phys. {\bf 96},  3866  (1992).

\bibitem{dai3}
J. Dai and J.~Z.~H. Zhang, J. Chem. Phys. {\bf 102},  6280  (1995).

\bibitem{tal-ezer1}
H. Tal-Ezer and R. Kosloff, J. Chem. Phys. {\bf 81},  3967  (1984).

\bibitem{kosloff2}
D. Kosloff and R. Kosloff, J. Comp. Phys. {\bf 52},  35  (1983).

\bibitem{pekeris1}
C.~L. Pekeris, Phys. Rev. {\bf 45},  98  (1934).

\bibitem{balint-kurti1}
G.~G. Balint-Kurti, R.~N. Dixon, and C.~C. Marston, J. Chem. Soc. Faraday
  Trans. {\bf 86},  1741  (1990).

\bibitem{balint-kurti2}
G.~G. Balint-Kurti, R.~N. Dixon, and C.~C. Marston, Int. Rev. Phys. Chem. {\bf
  11},  317  (1992).

\bibitem{mowrey2}
R.~C. Mowrey and G.~J. Kroes, J. Chem. Phys. {\bf 103},  1216  (1995).

\bibitem{neuhauser1}
D. Neuhauser and M. Baer, J. Chem. Phys. {\bf 90},  4351  (1989).

\bibitem{vibok1}
{\'{A}}. Vib{\'{o}}k and G.~G. Balint-Kurti, J. Chem. Phys. {\bf 96},  7615
  (1992).

\bibitem{vibok2}
{\'{A}}. Vib{\'{o}}k and G.~G. Balint-Kurti, J. Phys. Chem. {\bf 96},  8712
  (1992).

\end{thebibliography}

\end{document}